\documentclass[sigconf]{acmart}

\copyrightyear{2025}
\acmYear{2025}
\setcopyright{cc}
\setcctype{by}
\acmConference[CODASPY '25]{Proceedings of the Fifteenth ACM Conference on Data and Application Security and Privacy}{June 4--6, 2025}{Pittsburgh, PA, USA}
\acmBooktitle{Proceedings of the Fifteenth ACM Conference on Data and Application Security and Privacy (CODASPY '25), June 4--6, 2025, Pittsburgh, PA, USA}
\acmDOI{10.1145/3714393.3726006}
\acmISBN{979-8-4007-1476-4/2025/06}

\usepackage{hyperref}
\usepackage{amsfonts}
\usepackage{algorithmic}
\usepackage{graphicx}
\usepackage{textcomp}
\usepackage{booktabs}
\usepackage[nolist]{acronym}

\acrodef{dos}[DoS]{denial-of-service}
\acrodef{dss}[DSS]{distribution substation}
\acrodef{edf}[EDF]{environment description file}
\acrodef{fdi}[FDI]{false data injection}
\acrodef{ics}[ICS]{Industrial Control System}
\acrodef{ict}[ICT]{Information and Communication Technologies}
\acrodef{iids}[IIDS]{Industrial Intrusion Detection System}
\acrodef{ied}[IED]{Intelligent Electronic Device}
\acrodef{mitm}[MitM]{machine-in-the-middle}
\acrodef{rtu}[RTU]{remote terminal unit}
\acrodef{mtu}[MTU]{master terminal unit}
\acrodef{sdf}[SDF]{scenario description file}
\acrodef{sut}[SUT]{system under test}
\acrodef{swat}[SWaT]{Secure Water Treatment}
\acrodef{tss}[TSS]{transforming substation}
\acrodef{ot}[OT]{operational technology}

\usepackage{xcolor}
\usepackage[detect-weight=true,detect-family=true]{siunitx}
\usepackage{xspace}

\usepackage{amsmath}

\usepackage{tikz}
\usetikzlibrary{positioning}

\usepackage{multirow}

\usepackage{textgreek}

\definecolor{lightblue}{HTML}{2596be}
\definecolor{darkgrey}{RGB}{80,80,80}
\definecolor{lightgrey}{RGB}{170,170,170}

\definecolor{brown}{HTML}{a52a2a}
\definecolor{darkcyan}{HTML}{0a888a}

\usepackage{adjustbox}
\usepackage{booktabs}
\usepackage{wasysym}
\usepackage{tabularx} 
\usepackage{multirow}
\usepackage{arydshln} 
\usepackage{siunitx}

\newcommand\Tstrut{\rule{0pt}{6pt}}         
\newcommand\Bstrut{\rule{0pt}{0pt}}

\newcommand{\etal}{\textit{et~al.}\xspace}
\newcommand{\ie}{\textit{i.e.},\xspace}
\newcommand{\eg}{\textit{e.g.},\xspace}

\newcommand{\CUT}[1]{{\iffalse#1\fi}}

\newcommand{\wattson}{\textsc{Wattson}\xspace}

\newcommand{\name}{\textsc{Sherlock}\xspace}
\newcommand{\link}{\url{https://sherlock.wattson.it/}\xspace}
\newcommand{\cigrescenario}{\texttt{01-Basic}\xspace}
\newcommand{\semiurbscenario}{\texttt{02-Semiurban}\xspace}
\newcommand{\ruralscenario}{\texttt{03-Rural}\xspace}

\usepackage[most]{tcolorbox}
\newtcolorbox{challengebox}[1][]{
  enhanced,
  breakable,
  arc=0pt,
  outer arc = 0pt,
  sharp corners,
  boxrule=2pt,boxsep=0pt,top=2pt,left=2pt,right=2pt,bottom=2pt,middle=-5pt,
  colback=gray!10, 
  colframe=gray!10, 
  pad at break=0pt,bottomrule at break=0pt,toprule at break=0pt,
  borderline east={0.8pt}{-0.25pt}{black,dotted},
  borderline west={0.8pt}{-0.25pt}{black,dotted},
  borderline south={0.8pt}{-0.25pt}{black,dotted},
  borderline north={0.8pt}{-0.25pt}{black,dotted},
  title={\textcolor{black}{\textbf{#1}}},
}

\begin{document}

\title{Sherlock: A Dataset for Process-aware Intrusion Detection Research on Power Grid Networks}

\subtitle{Dataset Paper}

\author{Eric Wagner}
\email{eric.wagner@fkie.fraunhofer.de}
\orcid{0000-0003-3211-1015}
\authornote{Both authors contributed equally to this work}
\affiliation{%
  \institution{Fraunhofer FKIE\\RWTH Aachen University}
  \city{Aachen}
  \country{Germany}
}

\author{Lennart Bader}
\email{lennart.bader@fkie.fraunhofer.de}
\orcid{0000-0001-8549-1344}
\authornotemark[1]
\affiliation{%
  \institution{Fraunhofer FKIE\\RWTH Aachen University}
  \city{Aachen}
  \country{Germany}
}

\author{Konrad Wolsing}
\email{konrad.wolsing@fkie.fraunhofer.de}
\orcid{0000-0002-7571-0555}
\affiliation{%
  \institution{Fraunhofer FKIE\\RWTH Aachen University}
  \city{Aachen}
  \country{Germany}
}

\author{Martin Serror}
\email{martin.serror@fkie.fraunhofer.de}
\orcid{0000-0002-6925-5744}
\affiliation{%
  \institution{Fraunhofer FKIE}
  \city{Aachen}
  \country{Germany}
}

\begin{abstract}

Physically distributed components and legacy protocols make the protection of power grids against increasing cyberattack threats challenging.
Infamously, the 2015 and 2016 blackouts in Ukraine were caused by cyberattacks, and the German Federal Office for Information Security (BSI) recorded over 200 cyber incidents against the German energy sector between 2023 and 2024.
Intrusion detection promises to quickly detect such attacks and mitigate the worst consequences.
However, public datasets of realistic scenarios are vital to evaluate these systems.
This paper introduces \name, a dataset generated with the co-simulator \wattson.
In total, \name covers three scenarios with various attacks manipulating the process state by injecting malicious commands or manipulating measurement values.
We additionally test five recently-published intrusion detection systems on \name, highlighting specific challenges for intrusion detection in power grids.
Dataset and documentation are available at \link.

\end{abstract}

\ccsdesc[500]{Security and privacy~Intrusion detection systems}

\keywords{dataset, critical infrastructure, power grid, IEC 60870-5-104}

\maketitle

\section{Introduction}
\label{sec:introduction}

Cyberattacks against critical infrastructures, such as power grids, are on the rise~\cite{2020_alladi_industrial,2023_rajkumar_cyber}.
These attacks typically exploit vulnerabilities in the underlying \ac{ics} networks, which are known to rely on insecure legacy communication protocols~\cite{2012_galloway_introduction,SHH+21}. 
To make matters worse, such legacy protocols are difficult to replace due to long lifecycles of industrial hardware, stringent availability requirements, and limited update capabilities.
Under these circumstances, feasible preventive security measures, such as network segmentation and firewalls, do not suffice for protection~\cite{beaudet2020process}.
Effective intrusion detection promises to identify cyberattacks in their early stages and thus enable timely countermeasures that prevent severe damage.
\acp{iids} are thus widely recognized as a retrofittable, non-disruptive, and promising security solution, serving as the last line of defense for critical infrastructure~\cite{2023_lamberts_sok}.
However, such \acp{iids} must deal with the unique characteristics of power grid networks.

Traditional network-based \acp{iids} detect suspicious activities by scanning  traffic for known attack patterns~\cite{2022_kus_revisiting}.
However, \ac{ics} networks are often exposed to unknown attacks due to their distinct characteristics and the use of a wide range of different protocols.
Meanwhile, the predictable nature of control and sensor traffic, closely tied to underlying physical processes, creates opportunities for process-aware intrusion detection~\cite{beaudet2020process}.
Here, the core idea is to examine a system's physical state using data transmitted over the network to detect anomalies.
However, this approach necessitates domain-specific training data as a prerequisite for accurately modeling the expected system behavior.

This relatively new research area is experiencing significant growth, with at least 130 new process-aware intrusion detection mechanisms proposed in 2021 alone~\cite{2023_lamberts_sok}.
However, progress is hindered by a lack of high-quality datasets: fewer than half of publications use public datasets, and only \SI{16.4}{\%} utilize more than one.
Within critical infrastructures, the energy sector faces a notable dataset gap, as most existing datasets primarily represent small-scale networks or individual components~\cite{zemanek2022powerduck, AhKa21, 2019_biswas_synthesized}. 
Consequently, no comprehensive dataset is currently available for evaluating process-aware intrusion detection in power grid networks.

At the same time, the successful attacks on the Ukrainian power grid in 2015 and 2016, as well as the attempted attack in 2022, underscore the devastating impact of such incidents and highlight the interest of powerful adversaries exploiting these weaknesses~\cite{2017_whitehead_ukraine}.
Meanwhile, the German Federal Office for Information Security (BSI) reported over 200 cyber incidents against the Germany energy sector between 2023 and 2024~\cite{2023_bsi_report,2024_bsi_report}.

With this paper, we introduce \name{}, a comprehensive dataset specifically designed for process-aware, as well as network-based, intrusion detection.
\name{} was recorded using the co-simulator \wattson~\cite{2023_bader_wattson}, which can simulate power grids while concurrently emulating their corresponding \ac{ict} networks.
\wattson has been validated against a physical power grid, ensuring its accuracy, and supports both emulated network components and hardware-in-the-loop integration. 
Furthermore, it provides a safe research environment for replicating both routine operations and real-world cyberattacks.
These features make it an ideal tool for collecting representative, diverse, and reproducible datasets for intrusion detection research.

We passively capture network traffic from critical vantage points across three scenarios (denoted as \cigrescenario{}, \semiurbscenario{}, and \ruralscenario{}) during simulations spanning 35 days.
For two networks, we provide labeled attack-free and attack datasets, while for the third, only attack data is available.
This approach promotes research into the generalizability and transferability of detection methods, a critical goal given the expense and complexity of retraining intrusion detection systems with clean training data for each network.

Captured traffic is post-processed using the IPAL toolset~\cite{2022_wolsing_ipal} to generate time-series data representing the system's physical state.
This abstraction layer enhances dataset accessibility and decouples it from region-specific communication protocols.

The IPAL representation enables us to evaluate the detection performance of five \acp{iids}~\cite{2022_wolsing_simple, 2019_feng_systematic, 2020_kim_seq2seq, 2018_aoudi_pasad,2025_wolsing_geco}, which claim domain generalizability. 
This evaluation reveals six challenges in power grid networks that current process-aware intrusion detection mechanisms struggle to address, ranging from handling thousands of data points to accommodating long-term process variations.

\section{Related Work}


Conti~\etal~\cite{2021_conti_survey} identifies $23$ public datasets for cybersecurity research in \ac{ics} networks.
However, most lack process data and some rely on \acp{iids} explicitly learning from captured attack samples, which limits their ability to detect novel attacks or variations~\cite{2022_kus_revisiting}.
Among these, only five datasets of decent quality incorporate process data, yet none represent power grid scenarios at realistic scales.

The SWaT dataset~\cite{2016_goh_swat} captures process data from a scaled-down water treatment plant, featuring 36 physical attacks executed as \ac{mitm} attacks over 11 days.
Its single-execution attacks and focus on physical modifications limit its generalizability.
The WADI dataset~\cite{ahmed2017wadi} focuses on water distribution, incorporating 14 attacks over 16 days with more physical measurement points but fewer process stages than SWaT. 
Similarly, the BATADAL dataset~\cite{taormina2018battle}, based on a simulated water distribution network, samples data at hourly intervals, spanning a year with 14 attacks but offering coarser granularity.
The HAI dataset~\cite{SLYK20} covers different stages of power generation processes.
It features 50 diverse attacks of varying complexity. 
Finally, the EPIC dataset~\cite{AhKa21} focuses on a small power grid scenario with data collected from \acp{ied} monitoring electrical parameters. 
While it includes both network and process data, its attack scenarios are limited to malicious reconfigurations of devices.

Current datasets in the power grid domain~\cite{zemanek2022powerduck, AhKa21, 2019_biswas_synthesized} focus on network-based attacks and do either not include attacks against the physical process or do not even accurately reflect the physical process.
Thus, large-scale power grid scenarios are not covered by existing datasets.
Moreover, there remains a general lack of datasets that combine both network and process data, feature complex, multi-stage attacks, and are grounded in realistic, scalable scenarios.
\name should fill this gap.

\section{The \name Dataset}%
\label{sec:dataset}

The \name dataset aims to address gaps left by existing datasets, providing a valuable resource for assessing intrusion detection methods in power grids.
Beyond this goal, \name is designed to support broader research into power grid cybersecurity and the practical deployment of \acp{iids} in realistic network environments.
The following sections detail our testbed setup and provide an in-depth overview of the \name dataset. Additional information is available on the \name website at \link.

\subsection{Testbed Setup}

The backbone of our testbed is the \wattson simulator~\cite{2023_bader_wattson}. 
\wattson is an open-source power grid co-simulator, \ie it emulates realistic network traffic among power grid devices while simulating the power grid.
\wattson uses PowerOwl~\cite{powerowl} to model the power grid, which offers steady-state power flow calculations based on pandapower~\cite{2018_thurner_pandapower} and emulates its communication network supporting switches, routers, and hosts with lightweight namespaces, based on Docker containers, and with virtual machines in Linux. 
\wattson uses \texttt{tc} for traffic control to configure delays, jitter, bandwidth, and packet-loss for each individual link.
The communication between the control center and substations is performed using the IEC 60870-5-104 (IEC 104) protocol.

For our scenarios, we focus on future-oriented settings with a significant fraction of substations being digitized, \ie they digitally transmit measurements and---if applicable---support the remote execution of control commands.
For each scenario, we include load and optional generation profiles to control the behavior of these assets during the evaluation.
\wattson performs a real-time co-simulation with a 14x accelerated power profile, \ie evaluating \SI{12}{\hour} of network traffic reflects the power generation and usage patterns over an entire week.
To reduce complexity, we abstract from protective relays as intrusion detection should alert before they trigger.
The \name dataset is composed of network captures from mirror ports at switches that were identified as key vantage points, enriched with logs from individual hosts and services, additional context information, process ground-truth information, control center events, and documentation.
Our online documentation presents the details of the different scenarios, vantage points, and grid values.

For \name, we extract all relevant information passively, aligning with a non-invasive deployment strategy well-suited for real-world power grid networks.
The alternative of active polling consumes substantial bandwidth and exposes the \ac{iids} to manipulated data. 
Given that the centralized control center typically serves as a data sink, it provides an ideal vantage point, offering a comprehensive overview of the network. 
\name also provides data from alternative vantage points for further insights when desired.

In total, we simulate 35 days of power grid behavior, split into training and test sets of three different scenarios for \name.

\subsection{Scenarios}
\name{} contains three different scenarios of different size and complexity, each consisting of a power grid topology, an \ac{ict} network topology, and configurations regarding the coupling of both domains, \ie{} responsibilities of \acp{rtu} along with communication protocol information.
For each power grid topology, we use PowerOwl \cite{powerowl} to automatically detect facilities and derive a realistic \ac{ict} network, resulting in a simulation scenario compatible with \wattson.
The power grid itself comprises multiple stations, \ie{} several \emph{\acp{dss}} and one \emph{\ac{tss}}.
Each station contains one or multiple buses that are connected by lines and transformers and further link with assets such as storages (batteries), generators, and loads.

The \ac{ict} network comprises multiple subnets, with each scenario including at least two OT subnets for RTUs: one for the \ac{tss} and one or more for the \acp{dss}. 
Additionally, there is a Control Center subnet hosting the \ac{mtu} and multiple office subnets for servers and workstations. 
The subnets are interconnected via routers using the OSPF protocol and further include switches to link individual facilities and multiple hosts within these facilities.
The topologies are part of \name{}'s documentation and are explained further on the dataset website. 
Beyond the topologies, each scenario specifies the communication behavior of key assets, such as the \ac{mtu} and \acp{rtu}, as well as the operational behavior of power grid components like loads, storage systems, and generators.

For the IEC~104-based communication between the control center and the \acp{rtu}, we define an interval of \SI{10}{\second} for periodic measurement transmissions such that each \ac{rtu} transmits measured voltages, currents, and power values unsolicitedly.
Discrete values, such as binary states of circuit breakers, are only transmitted when explicitly requested and every time the value changes.

The power grid behavior is determined by the pre-defined behavior of loads, generators, and storages, further influenced by control operations executed by the grid operator.
These operations involve sending control commands to \acp{rtu}.
Depending on the individual scenario, the respective power profiles target all assets and vary across asset types.
For instance, a load representing a household exhibits different behavior compared to that of a supermarket. 
Whenever possible, we utilize profiles provided by the power grid scenarios; otherwise, we rely on a generic load curve as a fallback.

Next, we briefly introduce the three different scenarios featured in the \name{} dataset.

\subsubsection{\cigrescenario{}: The Cigre MV Reference Grid}
This scenario comprises \num{12} medium voltage (MV) \acp{dss} connected to a high voltage to medium voltage (HV/MV) \ac{tss}.
It includes \num{13} generators, \num{2} storages, \num{18} loads, and \num{2} HV/MV transformers.
With \num{32} \acp{rtu} distributed across two \ac{ot} subnets, each substation supports remote monitoring and control via a single \ac{mtu}.
We apply a generic load profile to all \num{18} loads, while storages and generators operate with static power infeed or consumption.
The scenario adopts the \emph{Cigre MV} power grid topology provided by pandapower~\cite{2018_thurner_pandapower}, based on the CIGRE Task Force C6.04.02 paper~\cite{strunz2005cigre}.

\subsubsection{\semiurbscenario{}: Simbench MV Semi-urban}
Complementing the \cigrescenario{} scenario, the \name{} dataset incorporates two larger, more realistic scenarios derived from Simbench~\cite{meinecke2020simbench}.
The \emph{Simbench MV Semi-urban} models an HV/MV distribution grid supplying a semi-urban city area. 
It features a central \ac{tss} with two transformers connecting to two double-busbars. 
Its \num{118} \acp{dss} follow a multi-ring topology and, like the Cigre MV scenario, connect a future-oriented number of renewable generators.
The \ac{ot} network includes \num{9} of \num{17} subnets, with all \num{72} \acp{rtu} linked to a single control center.
Unlike the Cigre MV scenario, this setup applies scenario-specific load and generation profiles to all relevant assets.
The power grid topology is based on the simbench key \texttt{1-MV-semiurb--2-sw}.

\subsubsection{\ruralscenario{}: Simbench MV Rural}
Transferability of \acp{iids} to similar yet different scenarios is a crucial research objective.
To support this, \name{} includes a third scenario that shares similarities with the \semiurbscenario{} scenario's topology but differs in size and asset count.
The power grid topology is based on the simbench key \texttt{1-MV-rural--2-sw} and compromises \num{95} \acp{dss} and a single \ac{tss}, representing a rural distribution grid.
The combined nominal power of all loads exceeds \SI{30}{{\mega\volt\ampere}}, while all generators provide \SI{47}{{\mega\volt\ampere}}.
With \num{12} \ac{ot} subnets, \num{16} subnets in total, and \num{60} \acp{rtu}, the \ac{ict} network is smaller compared to \semiurbscenario{}.
Providing no training data, the \name{} dataset encourages researchers to enhance transferability by training their \acp{iids} with the \semiurbscenario{} scenario and testing them against the \ruralscenario{} scenario.

\subsection{Commands and Measurements}
In all scenarios, \acp{rtu} monitor and control power grid assets, including buses, lines, transformers, circuit breakers, loads, generators, and batteries.
Most floating-point measurements that are expected to change gradually, such as voltages and currents, are periodically transmitted to the control center using the IEC~104 protocol (\texttt{Type ID=13}, \texttt{Cause of Transmission (CoT)=1}).
Other data points, such as booleans (\texttt{Type ID=1}), are configured to be transmitted spontaneously (\texttt{CoT=3}), \ie{} when they are changed.
This includes tap positions on transformers, circuit breaker states, and the connectivity of loads, generators, and storages.

For control commands, the \ac{mtu} issues target values to the \acp{rtu} with desired states, \eg{} for circuit breaker positions or power infeed set points.
These commands are verified by the responsible \ac{rtu}, executed, and an acknowledgment is sent back to the \ac{mtu}.
In case of invalid or unrealizable commands, a negative confirmation is sent.
During normal operation, the control center can reduce and increase the power infeed of generators or change the topology---either to reduce the load on transformers and lines or to allow maintenance work in distribution substations or on power lines.
For the dataset, benign commands issued by the power grid operator will not impede the power supply for customers.

\subsection{Attacks}

\begin{table}
  \centering \footnotesize
  \setlength\abovecaptionskip{-10pt}
  \newcolumntype{Z}[0]{>{\Tstrut\centering}X<{\Bstrut}}

  \begin{tabularx}{\columnwidth}{cZcZZZ}
     & & \textbf{Vantage } & & & \textbf{Benign} \tabularnewline
    \textbf{Scenario} & \textbf{Type} & \textbf{Points} & \textbf{Duration} & \textbf{Attacks} & \textbf{Events} \tabularnewline\midrule
    \multirow{2}{*}{\cigrescenario{}} & train & 4 & 12 h & - & 7 \tabularnewline \cline{2-6} 
    & test & 4 & 12 h & 17 & 10
    \tabularnewline \midrule
    \multirow{2}{*}{\semiurbscenario{}} & train & 6 & 12 h & - & 9 \tabularnewline \cline{2-6} 
    & test & 6 & 12 h & 29 & 10
    \tabularnewline \midrule
    \ruralscenario{} & test & 8 & 12 h & 28 & 8
    \tabularnewline \midrule
  \end{tabularx}
  \caption{Metadata of \name's scenarios.}
  \label{tab:scenarios}
\end{table}

There exist many paths for an attacker to get to a state where they have full control over one or multiple devices in a network.
These approaches range from supply chain attacks over physical intrusions, such as breaching a substation to connect unauthorized devices, to the classic exploitation of vulnerabilities in existing devices.
Some of these steps are not detectable by \acp{iids}~(\eg supply chain attacks) and others are device-specific~(\eg exploitation of devices).
Therefore, we focus on the final phase of attacks that actively impact the state of the power grid, either by injecting control commands, suppressing messaging, or manipulating measurements to provoke damaging reactions or hide critical conditions.

Table~\ref{tab:scenarios} provides an overview of the three scenarios and their respective metadata. 
The attacks are executed consecutively within a single run, with clearly defined start and end points. 
Sufficient time is allocated between attacks to allow the system to recover to a stable state. 
Additionally, we include multiple extended attack-free periods in the test set to help minimize false alarms.

Table~\ref{tab:attacks} presents simple examples from the \cigrescenario{} scenario of the four attack types covered by \name. 
These attack types include \ac{dos}, control command insertion inspired by the Industroyer attack that caused widespread blackouts in the Ukrainian power grid~\cite{industroyer2blogpost}, and advanced false data injection attacks that distort the grid operator's view of the grid state. 
The \emph{Control \& Freeze} attack further manipulates the grid in real-time. 
Detailed descriptions of all attacks, as well as maintenance events that may be mistakenly identified as attacks, are provided in the documentation.

\subsection{Recommended Evaluation Metrics}

We recommend that researchers utilizing \name for evaluating their \ac{iids} primarily report three metrics: Detected Attacks, False Alarms, and Average Time to Detection (TTD).
To calculate these metrics, we define an alarm as a continuous signalization of an attack by an IIDS.
These metrics are then defined as follows:


\textbf{Detected Attacks.}
The absolute number of attacks during which an alarm starts. Alarms starting before the attack are considered false alarms. Alarms may start during a recovery phase while the system returns to normal operation. Such alarms should be ignored and neither count as detected attack nor as false alarm.

\textbf{False Alarms.}
The number of alarms that start during normal operations without an attack. Maintenance events should also be considered normal operation. One may indicate how many of the false alarms are triggered by such events.

\textbf{Average TTD.}
The average time in seconds from the start of an attack until the first alarm starts.

Incorporating additional time-based metrics, such as eTa~\cite{2022_hwang_eta}, and performance indicators, such as training and classification time, provide deeper insights into system capabilities. 
Note that point-based metrics, such as F1-scores, are generally unsuitable for assessing time-aware IDSs~\cite{2022_hwang_eta} since such metrics insufficiently penalize false alarms and fail to account for scenarios where a brief alarm at the start of an attack yields artificially high scores.

\begin{table}
  \centering \footnotesize
  \setlength\abovecaptionskip{-10pt}
  \begin{tabularx}{\columnwidth}{p{1cm}p{8mm}p{5.7cm}}
    \textbf{Attack Type} & \textbf{Start}\newline\textbf{End} & \textbf{Description} \tabularnewline\midrule
    DoS & 04:11:23 \newline 04:13:30 & ARP spoofing attack against \acp{rtu} \num{127} and \num{128}, interrupting the \ac{mtu} connection.
    \tabularnewline \midrule
    Industroyer & 04:55:28 \newline 04:58:34 & A secondary IEC~104 client connects to \ac{rtu}~\num{123} from compromized \ac{rtu}~\num{121} and issues control commands every \SI{3}{\second} to disconnect circuit breakers \num{15} and \num{16}, inducing a blackout at \ac{dss} \num{8} (Bus~\num{5}).
    \tabularnewline \midrule
    Drift Off & 07:56:27 \newline 08:06:26 & As \ac{mitm} between the \ac{mtu} and \ac{rtu}~\num{118}, the attacker manipulates the voltage measurements regarding Bus~\num{2} at \ac{dss}~\num{5} to gradually increase to \num{1.37} (\SI{27}{\kilo\volt}).
    \tabularnewline \midrule
    Control \& Freeze & 10:04:27 \newline 10:13:45 & As \ac{mitm}, the attacker learns measurement trends from multiple \acp{rtu} and continues to manipulate future measurements to match this trend after injecting control commands that gradually reduce the power infeed of Generator~\num{7}, masking the command's local effects.\tabularnewline \midrule
  \end{tabularx}
  \caption{Examples of the different attack types on \cigrescenario{}.}
  \label{tab:attacks}
\end{table}

\subsection{Data Format and Extraction}

The \name dataset primarily encompasses packet capture in three scenarios from different vantage points.
Additionally, \name provides device logs, context information about events (maintenance and attacks), data point mappings to human-readable identifiers, and ground truth information about the power grid state.
To enhance accessibility for researchers, we additionally convert the dataset into the IPAL~\cite{2022_wolsing_ipal} format, an abstract representation of network data specifically designed for intrusion detection research.

Primarily, we focus on passively extracting the current state of the power grid from network traffic.
Therefore, we log the initial system state and parse each intercepted IEC 104 packet to update this system state while mapping abstract Information Object Addresses~(IOAs) to human-readable identifiers.
This observed state, recorded from a single vantage point, is logged every second.
As a result, detection performance on the \name dataset reflects what would be achieved in a passive, real-world deployment.

\section{Challenges of IDSs in Power Grids}

Providing \name in the IPAL format makes it more accessible to the research community.
Additionally, it enables us to benchmark a range of existing \acp{iids} adapted to IPAL that promise domain-independent industrial intrusion detection.
We tested five \acp{iids} on the \name dataset and present the results in Sec.~\ref{sec:challenges:results}:


\textbf{PASAD}~\cite{2018_aoudi_pasad} interprets a process value as a vector space and assesses the drift compared to the behavior observed during training.

\textbf{Invariant}~\cite{2019_feng_systematic} automatically learns invariants of a process that should hold at all times and alerts if any invariant is violated.

\textbf{Seq2SeqNN}~\cite{2020_kim_seq2seq} is a neural network trained to predict the next process state that alerts significant deviations from the prediction.

\textbf{SIMPLE}~\cite{2022_wolsing_simple} assesses process values' plausibility based on extrema, changes, and distributions observed during training.

\textbf{GeCo}~\cite{2025_wolsing_geco} learns a state-space model of the entire process and alerts upon deviations from predicted behavior.

Based on these results, we then identify the specific challenges of intrusion detection in the power grid domain in Sec.~\ref{sec:challenges:challenges}.

\subsection{Evaluation Results}
\label{sec:challenges:results}

\begin{figure}
  \centering
  \setlength\abovecaptionskip{-3pt}
  \setlength\belowcaptionskip{-10pt}
  \includegraphics[width=\columnwidth]{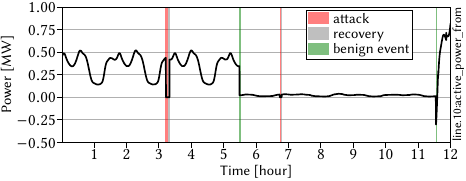}
  \caption{
    Time-series data shows outliers during attacks, but benign switching event also drastically change the behavior over prolonged times.
  }
  \label{fig:data}
\end{figure}

In total, we trained five \acp{iids} for the relatively small \cigrescenario scenario, which features few stations and data points compared to more realistic networks such as the \semiurbscenario and \ruralscenario scenarios.
Table~\ref{tab:ids} presents our evaluation results.
Of these \acp{iids}, PASAD and Invariant did not produce useful results. Instead, they generated one continuous alert for most of the test data. 

The other three \acp{iids} also only achieve mediocre attack detection performance.
A key challenge lies in the dynamic nature of power grids, which exhibit constantly shifting stable configurations driven by factors such as maintenance activities, energy generation and consumption patterns, and the state of connected superordinate grids.
Figure~\ref{fig:data} clearly illustrates this phenomenon.
Two attacks lead to clear outliers in the measured power.
However, the two highlighted switching events drastically change the power lines' characteristics, such that trivial intrusion detection mechanisms have a hard time identifying attacks.

To still investigate the potential of process-aware intrusion detection in power grids, we filtered data points that substantially changed between the test and training phases.
While this filtering approach facilitates assessing process-aware IDS performance under idealized conditions, it is not feasible in real-world deployments where all measurements may eventually be affected by such changes in the power grid.
Nevertheless, this approach provides insight into how process-aware IDS could perform if supplemented with contextual information, such as schedules for planned maintenance activities.

After filtering, Seq2SeqNN, SIMPLE, and GeCo show strong detection performance.
An effective \ac{iids} should produce near-zero false alarms, detect most attacks, and do so with minimal delay.
While process-aware intrusion detection shows promising potential, the power grid domain still faces several unresolved challenges.

\subsection{Challenges in Power Grids}
\label{sec:challenges:challenges}

Our evaluation shows the potential of process-aware intrusion detection for power grids.
The achieved detection performance will likely improve with further research into detection methodologies facilitated by \name.
However, these results were achieved only after filtering sensor values impacted by maintenance activities or switching operations. 
Our initial benchmarking highlights six challenges in implementing process-aware intrusion detection in power grids, which also exist in other domains to varying degrees.

Some \acp{iids} faced difficulties during training even on the small-scale \cigrescenario reference grid, despite its relatively limited number of substations.
Furthermore, each substation in \name transmits a minimal amount of data as we employ a single-phase power grid model and, due to the steady-state simulation, abstract from features such as the power grid's utility frequency.
\acp{iids} designed for power grids may reach scalability limitations in practical deployments, a shared problem with other industrial domains~\cite{2017_etalle_From}.

\begin{table}
  \centering \footnotesize
  \setlength\abovecaptionskip{-10pt}

  \begin{tabularx}{\columnwidth}{lcccc}
    \multicolumn{1}{c}{\textbf{\ac{iids}}} & \textbf{Detected Attacks} & \textbf{False Alarms*} & \textbf{Average TTD} \tabularnewline\midrule
    PASAD~\cite{2018_aoudi_pasad} & 0/18 & 1(0) & -- \tabularnewline \midrule
    Invariant~\cite{2019_feng_systematic} & 0/18 & 1(0) & --  \tabularnewline \midrule
    Seq2SeqNN~\cite{2020_kim_seq2seq} & 7/18 & 10(0) & 149.63s \tabularnewline \midrule
    SIMPLE~\cite{2022_wolsing_simple} & 16/18 & 33(33) & 81.43s \tabularnewline \midrule
    GeCo~\cite{2025_wolsing_geco} & 15/18 & 94(30) & 75.89s  \tabularnewline \midrule
    \multicolumn{4}{r}{* (x) shows how many of these alarms are cause by benign events}
  \end{tabularx}
  \caption{Detection performance of select \ac{iids} in the \textit{Cigre MV} scenario after heavily filtering measurements that are affected by switching operations and maintenance.}
  \label{tab:ids}
\end{table}

\begin{challengebox}[Challenge 1 -- Scalability]

  Intrusion detection must be capable of handling the frequent transmission of hundreds to thousands of data points that are generated by power grid operations.

\end{challengebox}

Beyond the sheer volume of training data, it is impossible to observe all possible states during the training. 
\acp{iids} typically aim to learn the cyclic and repetitive behavior of cyber-physical processes in order to detect anomalies.
Ideally, the training phase would span multiple cycles to capture this behavior.
However, in power grids, this cyclic behavior is only partially present due to factors such as daily weather changes, seasonal variations, maintenance operations, multiple stable configurations, and the ongoing integration of new components (\eg wind turbines). 
As a result, some perfectly valid grid configurations may never be observed during training.

\begin{challengebox}[Challenge 2 -- Training Data Limitation]
  
  Even with attack-free training data collected over an extended period, some entirely valid grid configurations would likely remain unobserved.

\end{challengebox}

The operation of a power grid introduces regular benign anomalies, which pose an additional challenge for intrusion detection.
Maintenance operations, for example, often require switching off specific power lines and redirecting power flows.
Additionally, changes in power demand and generation, including those in subordinate power grids, as well as equipment failures, can necessitate adjustments to the power grid configuration to prevent overloading particular lines.
Ideally, these benign changes should not trigger alarms in an \ac{iids}, or at least not result in prolonged false alarms.

\begin{challengebox}[Challenge 3 -- Benign Anomalous Behavior]
  Benign anomalous behavior should be anticipated and not trigger (prolonged) false alarms, ensuring dependable surveillance.
\end{challengebox}


An additional challenge for intrusion detection in power grids is the need to consider multiple vantage points. 
While the control center acts as the primary sink for most measurements and the origin for control commands, a network capture taken just in front of it does not provide a complete picture. 
Some communications may be missed, network-based intrusion detection could be compromised, and a strategically positioned attacker could manipulate data at individual vantage points. 
To address these issues, intrusion detection systems should incorporate multiple vantage points while minimizing communication overhead between them.

\begin{challengebox}[Challenge 4 -- Vantage Points]
  To ensure reliable intrusion detection, multiple vantage points should be considered while minimizing the resulting communication overhead.
\end{challengebox}

Multiple vantage points can also help in localizing and understanding the origin of an anomaly. 
In addition to confirming that these effects are due to a cyberattack, \acp{iids} should ideally assist in localizing the attack's origin. 
While actionability is, in general, desirable for \acp{iids}~\cite{2017_etalle_From}, power grids span vast geographical areas, making it time-consuming to travel between substations and investigate potential signs of an attack (\eg compromised devices).

\begin{challengebox}[Challenge 5 -- Actionability]
  \acp{iids} should not only detect attacks but also aid in understanding and pinpointing their origin, facilitating a quick resolution.
\end{challengebox}

Finally, different intrusion detection mechanisms are required to detect attacks as reliably as possible~\cite{2023_lamberts_sok}. 
Process-aware intrusion detection can quickly identify anomalous behavior, even in the absence of changes in network traffic, such as when a supply chain attack compromises an \ac{rtu} that confirms but does not execute commands. 
In contrast, network-based mechanisms may detect the attachment of new devices to the network (\eg by observing unexpected ARP requests), thereby identifying attacks before they manipulate the process. 
Meanwhile, host-based mechanisms may detect manipulated firmware in advance.

\begin{challengebox}[Challenge 6 -- Multi-layer Intrusion Detection]

  A holistic surveillance of power grids is only achievable by combining process-aware, network-based, and host-based intrusion detection. These approaches should ideally complement and support each other to enhance detection performance.
\end{challengebox}

\section{Conclusion}
\label{sec:conclusion}

We present the \name dataset to advance research on process-aware intrusion detection in power grids.
The dataset encompasses three scenarios of realistically sized power grids, passively capturing network traffic at multiple vantage points during normal operations and periods influenced by cyberattacks.
We extract process state information using human-readable data point identifiers in the IPAL format.
This format also facilitates testing and validation, as demonstrated by our evaluation of five general-purpose industrial intrusion detection methods on the dataset.
Our initial findings identify six key challenges for intrusion detection research for power grids, such as overlapping cyclic behaviors based on time of day, season, and weather, which complicate the identification of benign characteristics.
We envision that the \name dataset will assist the research community in tackling these challenges in the future and thus contribute to the security of critical infrastructures.

\section*{Acknowledgments}
This paper was supported by the EDA Cyber R\&T
project ``Cyber Electromagnetic Resilience Evaluation on Replicated
Environment (CERERE)'', funded by Italy and Germany.

\vspace{-0.5em}
\bibliographystyle{ACM-Reference-Format}
\bibliography{paper}

\end{document}